\documentclass[
aps,prl,
superscriptaddress,
twocolumn,
10pt,
amsmath,amssymb
]{revtex4}
\usepackage{graphicx}
\usepackage{graphics}
\usepackage{dcolumn}
\usepackage{bm}

\begin{document}

\title{Optical extinction in a single layer of nanorods}

\author{Petru Ghenuche}
\altaffiliation{Permanent address: Institute for Space Sciences, P.O. Box MG-23, RO-077125 Bucharest-M\u agurele, Romania}
\affiliation{Laboratoire de Photonique et Nanostructures (LPN-CNRS), Route de Nozay, 91460 Marcoussis, France}
\affiliation{Onera - The French Aerospace Lab, Chemin de la Huni\`ere, 91761 Palaiseau, France}

\author{Gr\'egory Vincent}
\affiliation{Onera - The French Aerospace Lab, Chemin de la Huni\`ere, 91761 Palaiseau, France}

\author{Marine Laroche}
\affiliation{Laboratoire Charles Fabry, Institut d'Optique, CNRS, Universit\'e Paris-Sud, Campus Polytechnique, RD 128, 91127 Palaiseau cedex, France}

\author{Nathalie Bardou}
\affiliation{Laboratoire de Photonique et Nanostructures (LPN-CNRS), Route de Nozay, 91460 Marcoussis, France}

\author{Riad Ha\"\i{}dar}
\affiliation{Onera - The French Aerospace Lab, Chemin de la Huni\`ere, 91761 Palaiseau, France}

\author{Jean-Luc Pelouard}
\affiliation{Laboratoire de Photonique et Nanostructures (LPN-CNRS), Route de Nozay, 91460 Marcoussis, France}

\author{St\'ephane Collin}
\email{stephane.collin@lpn.cnrs.fr}
\affiliation{Laboratoire de Photonique et Nanostructures (LPN-CNRS), Route de Nozay, 91460 Marcoussis, France}

\date{\today}

\begin{abstract}
We demonstrate that almost 100~\% of incident photons can interact with a monolayer
of scatterers in a symmetrical environment.
Nearly-perfect optical extinction through free-standing transparent nanorod arrays
has been measured. The sharp spectral opacity window, in the form of a characteristic Fano resonance,
arises from the coherent multiple scattering in the array.
In addition, we show that nanorods made of absorbing material exhibit a 25-fold
absorption enhancement per unit volume compared to unstructured thin film.
These results open new perspectives for light management in high-Q, low volume
dielectric nanostructures, with potential applications in optical systems, spectroscopy, and optomechanics.
\end{abstract}

\pacs{dielectric membranes, optical properties, etc}

\maketitle

Enhancing light-matter interactions at the nanometer scale is a key for many
applications in the area of photonics, biophysics and material sciences \cite{Barnes2003,Novotny2006}. 
Metallic particles are the archetype of nano-objects that can lead to strong interaction
with light due to plasmonic resonances, with the drawback of intrinsic metal absorption \cite{Wegener08}.
Conversely, the weak scattering cross-section of tiny dielectric nanoparticles
originates from their non-resonant nature, making them inefficient for optical
manipulation at the nanoscale.
However, the extinction cross-section can be increased by coherent multiple scattering in assemblies
of nanoparticles, offering another degree of freedom to manipulate their
optical response. It has been shown theoretically that it can lead to extremely sharp geometric
resonances \cite{Schatz, Markel, Abajo_RMP07, Gomez_OE06, Laroche_PRB06}.
From the experimental point of view, arrays of resonant, metallic nanoparticles
demonstrated the potential of the effect \cite{Krenn_PRL00, Hicks_NanoLett05}, making the
localized surface plasmon resonance much narrower \cite{Grigorenko_PRL08,Barnes_PRL08}.
Arrays of non-resonant, dielectric nanorods should offer new possibilities
for even higher quality-factor geometric resonances \cite{Laroche_OL07, Pellegrini_ACS09}.
Indeed, sharp reflection resonances \cite{Gomez_OE06} and absorption enhancement \cite{Laroche_PRB06}
have been predicted for dielectric, sub-wavelength cylinder arrays. 

It has to be emphasized that the scattering properties of nanoparticles are substantially modified, or even suppressed, in the proximity of a surface \cite{BohrenHuffman,StuartHallPRL98,Gomez_OE06, Abajo_RMP07,Vecchi_PRL09,Auguie2010}.
Here, we study a free-standing array of nanorods, with a filling fraction around 0.15. In contrast with the periodic nanostructures studied extensively in the past ten years (nanohole arrays in metal films~\cite{EbessenNat98,LalanneNat08}, metal nanoparticle arrays~\cite{Krenn_PRL00, Hicks_NanoLett05,Grigorenko_PRL08,Barnes_PRL08}, high-index-contrast gratings~\cite{Huang2007}, guided-mode resonant structures~\cite{WangOE09,Sakat11},...), resonant effects can not be attributed to localized resonances neither to interactions between nanostructures mediated by electromagnetic waves in matter (surface waves or guided modes dielectric in layers).

A free-standing monolayer of dielectric nanorods can be considered as a model structure for direct interactions in free space between non-resonant structures. In this letter, geometric resonances are revealed by angle-resolved transmission and reflection measurements. We demonstrate nearly-perfect optical extinction, and more than one order of magnitude absorption enhancement as compared to a same-volume thin film. The results can be fully explained by a multiple scattering effect and are in good agreement with an analytical model which includes the geometry and the material properties of the rods. It is also shown numerically that very large electric-field intensities can be obtained at the resonance wavelength, with field enhancements as large as $10^5$ for nanorod diameters of 50 nm.



\begin{figure*}
\begin{center}
\includegraphics[width=\columnwidth]{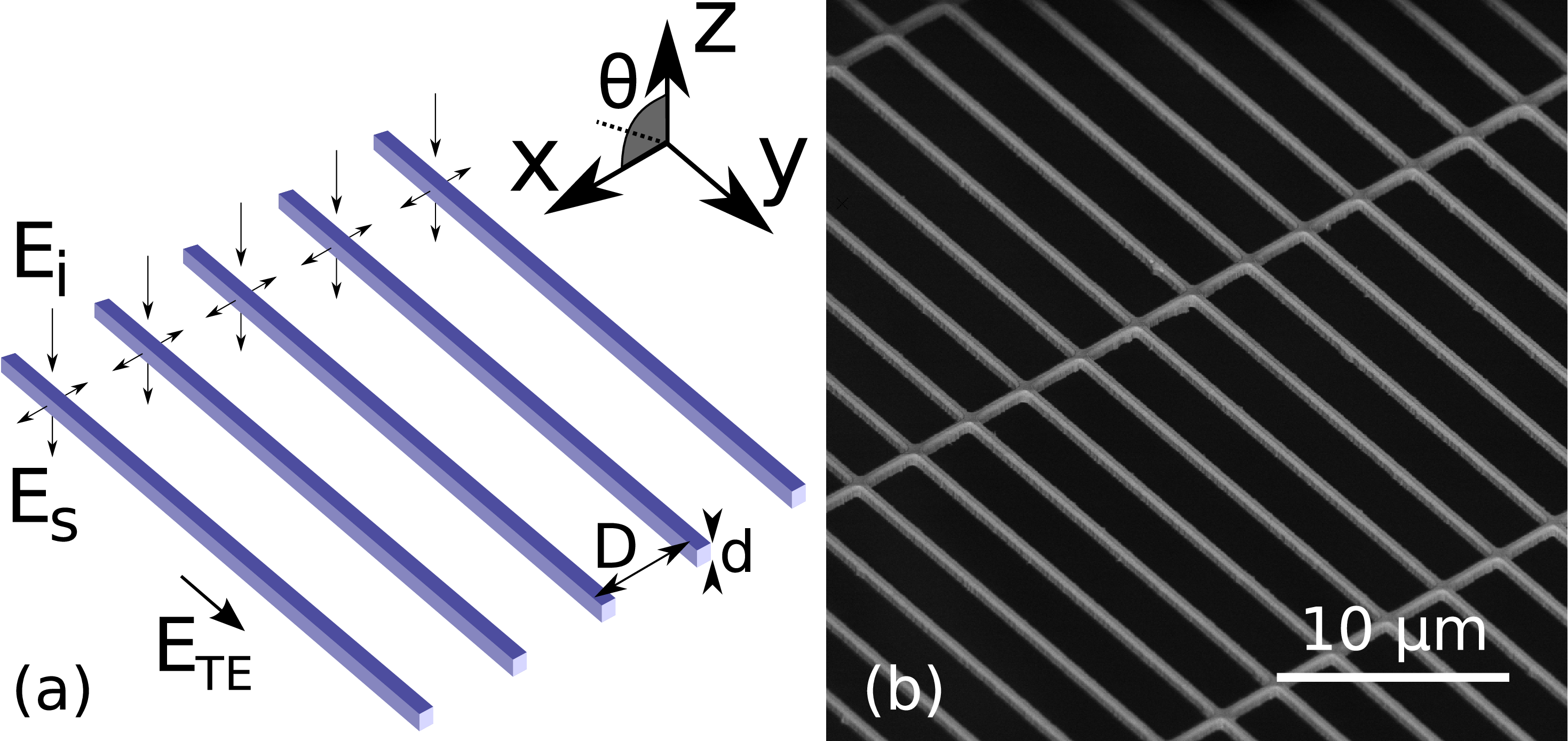}
\caption{
Free-standing dielectric grating. {\bf a}, The schematic illustrates the geometrical parameters of
the structure, measurement arrangement and the multiple scattering mechanism. {\bf b}, Scanning electron microscopy image
of a 1D free-standing ${\rm SiN}_{\rm x}$ membrane of $3 \: \rm \mu m$ period ($D$) made of square rods of $500$ nm size ($d$). 
} \label{photo_membrane}
\end{center}
\end{figure*}

Silicon nitride (${\rm SiN}_{\rm x}$) deposited by plasma-enhanced chemical vapor deposition (PECVD) exhibits both
transparent and absorbent optical behaviors for wavelengths between 1.5 - 5.5~$\mu$m \cite{Macek}.
For this reason, it has been used to fabricate the studied free-standing membranes made
of subwavelength rods. Membranes with 1D patterns were developed on Si
substrates and drilled by dry etching \cite{vincent:1852,Supl.Info}
on large surface areas (2.6$\times$2.6 mm$^2$), having 500 nm square section bars and 3 $\mu$m period.
To increase the rigidity of the membrane, support bars were
added in the fabrication design, as shown in Fig. \ref{photo_membrane}(b).
Their period was set to be large enough ($20 \: \mu$m) to ensure a negligible optical effect
in the spectral range of interest. 


Angle-resolved absolute transmission (T) and reflection (R) measurements were performed using
a Fourier-transform infrared spectrometer with an InSb detector.
The specular reflection measurements have been normalized with a gold mirror placed
in the proximity of the membrane. 
The spectral resolution was set to 5 cm$^{-1}$ (wavelength range: 1.5 - 5.5~$\mu$m) and a
home-made achromatic optical system allows a $\pm \ 0.5 ^{\circ}$ angular resolution \cite{Billaudeau:08}.
An artifact observed around 4.25~$\mu$m is produced by the change of the CO$_2$ content in air during the experiments.
The free-standing membrane is illuminated in the plane perpendicular to the rod axis with an
incident wave vector  $k = (k_x , k_z)$. The spectra have been recorded under an incidence angle $\theta$ ranging from $0^{\circ}$
to $60^{\circ}$ in $0.5^{\circ}$ increments. The illumination area was set to $2.2$~mm$^2$
to take into account a large number of rods (over 500).

\begin{figure*}
\begin{center}
\includegraphics[width=\columnwidth]{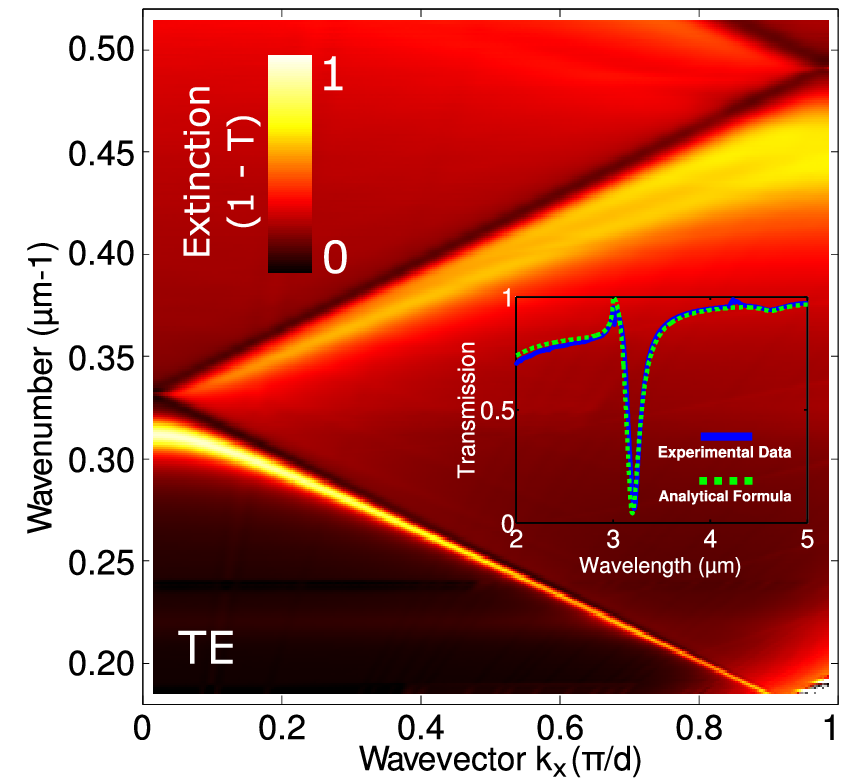}
\caption{
Extinction coefficient in the $(\omega,k)$ plane. Close to Rayleigh frequencies,
almost complete extinction is achieved due to geometric resonance. (Inset: Theoretical
and experimental transmission spectra at normal incidence, TE).} \label{ext_omega_k}
\end{center}
\end{figure*}

In Fig. \ref{ext_omega_k} is plotted the experimental dispersion diagram of the optical extinction 1-T$(\omega,k_x)$,
where T$(\omega,k_x)$ is the specular transmission in TE polarization (electric field parallel to the rod axis),
$\omega~=~2\pi c/\lambda$ is the frequency, $\lambda$ is the wavelength, and $k_x~=~(\omega/c) \sin \theta$ is
the $x$-component of the incident wave vector. This diagram shows sharp-bright features. At normal incidence
only one is visible, the corresponding transmission spectrum (inset, Fig. \ref{ext_omega_k}) reveals that
nearly total optical extinction (94~$\%$) is achieved in spite of (i)  the low  fill factor (15~$\%$) of the grating and (ii) the perfect
transparency of the material (no loss at 3.2~$\mu$m). The optical response of the dielectric membrane
changes in a very narrow spectral range from transparent to opaque behavior. From almost total transmission
at 3~$\mu$m, the signal drops sharply at 3.2~$\mu$m to just 6~$\%$.

The dispersion diagram is characterized by two branches with bright-dark bands. The resonances occur close to the
onset of a new propagating order, i.e close to the Rayleigh frequency given by
$ \lambda_m~=~D/m ( \pm 1 - \sin(\theta))$, where $m$ is an integer (m=$\pm 1$ in Fig. \ref{ext_omega_k}).
The optical transmission of this system can be considered as the sum of two components:
direct transmission and light scattered by the rods, inducing constructive
and destructive interferences with characteristic Fano lineshape \cite{Fanorew}. 

The physical mechanism of this geometric resonance is closely related to a multiple scattering mechanism.
Each subwavelength rod behaves like an individual scatterer with a dipolar response. It can be modeled as an
infinite circular cylinder with a polarizability
$\alpha~\approx~\alpha_0 \left[ 1 - i \frac{k^2}{4}\alpha_0 \right]^{-1}$ where
$\alpha_0~=~\pi  a^2(\epsilon-1)$ is the static polarizability, $a$ the cylinder radius and $\epsilon$
is the dielectric constant of the rods. When illuminated by a plane wave, each rod of the array
re-radiates a field proportional to its dipole moment \textbf{$p_y$}. Conversely, the incident field
on each rod is the sum of the incident plane wave (E$_i$) and the scattered fields (E$_s$)
from the other rods, as sketched in Fig. \ref{photo_membrane}(a).
The dipolar moment can be written as:
\begin{equation}
p_y =  \widehat{\alpha} \ E_i, \ \widehat{\alpha}  = \left( \frac{1}{\alpha} - G_b k^2 \right)^{-1},
\end{equation} 
where the effective polarizability $\widehat{\alpha}$ accounts for the multiple scattering mechanism.
$G_b$ is the dynamic depolarization term which only depends on the geometrical parameters,
 i.e. the period $D$, the radius $a$, as well as the wavelength and the angle of incidence $\theta$.
The geometric resonance is defined by the pole of the effective polarizability. 
This description leads to an analytical model for the far-field response of the
membrane \cite{Gomez_OE06}. 

To effectively compare the measured and calculated spectra of T and R, 
the parameters used in the model are the following:  grating constant $D$~=~3 $\mu$m, radius of the
rods $a$ = 0.3~$\mu$m to maintain the area equivalence of the rod section. In the spectral range of
interest, the dielectric constant of ${\rm SiN}_{\rm x}$ material is influenced by absorption
bands due to the stretching of N-H and Si-H bonds at about 3 and 4.6~$\mu$m, respectively.
These bounds are related with the presence of hydrogen in the PECVD deposition process \cite{Macek}. 
To account for these effects, the complex values of the dielectric constant have been retrieved by using the
reflection and transmission measurements of an unstructured ${\rm SiN}_{\rm x}$ membrane with similar
thickness and fabrication method \cite{Supl.Info}. At last, the spectral shift of the resonance within the convergence angle
of the incident beam is of the order of its linewidth. This effect has been taken into account by 
convoluting the calculated spectra with a 0.5~$^\circ$ waist Gaussian profile \cite{Supl.Info}.

\begin{figure*}
\begin{center}
\includegraphics[width=\columnwidth]{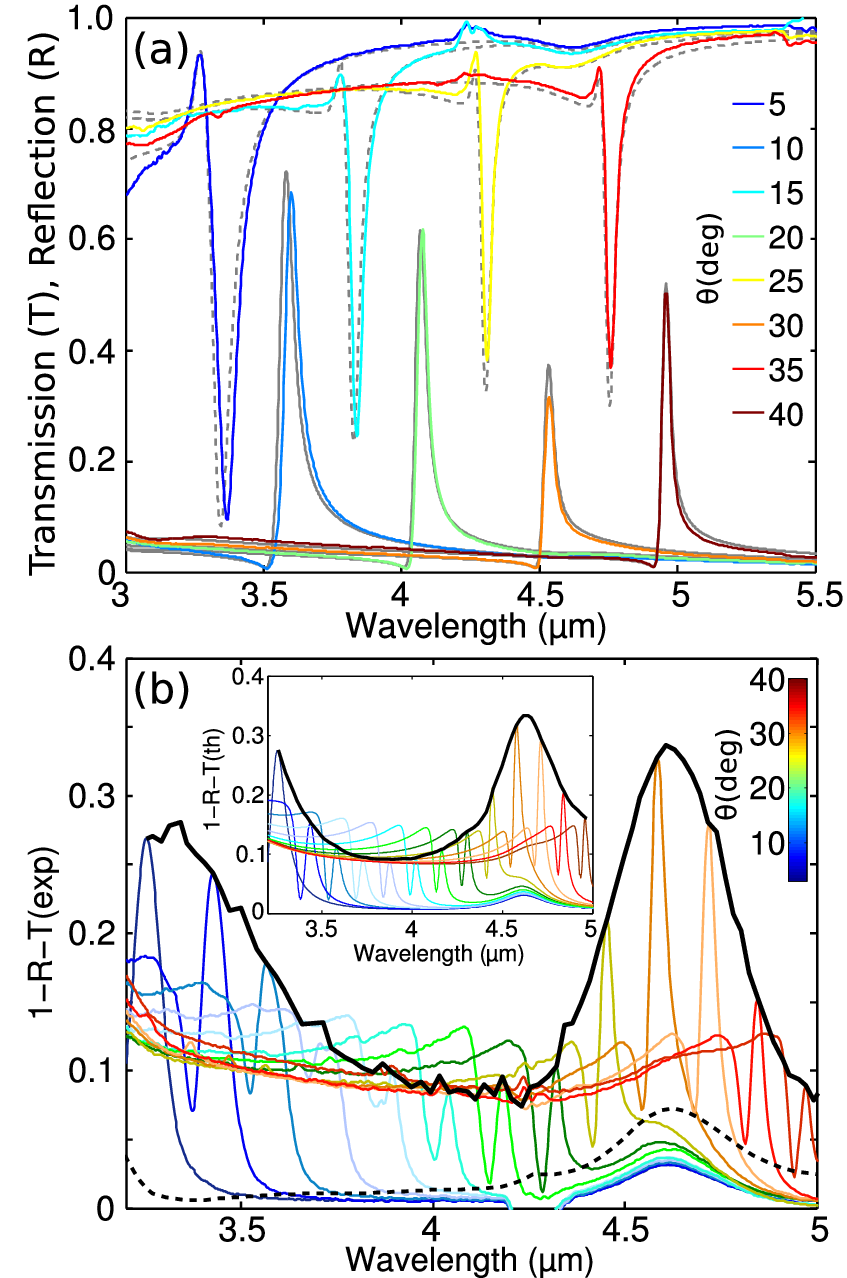}
\caption{
Experimental and theoretical spectra for different incidence angles.
{\bf a}, Transmission and reflection (simulated spectra: gray dotted lines).
{\bf b}, Absorption spectra and the absorption the unstructured membrane at $\theta=$~15$^\circ$ (black dotted line)
(inset: the corresponding simulated absorption spectra). 
The envelopes of the absorption maxima at resonance are depicted in black full lines.}
\label{comp_expe_analytical}
\end{center}
\end{figure*}

The results are shown in Fig. \ref{comp_expe_analytical}(a) for eight different angles of incidence
($5$ to $40$ degrees, alternated for clarity, TE polarization). A quantitative agreement is observed
between the experimental and the theoretical results for both transmission and reflection.
The small remaining discrepancies between simulated and experimental curves are attributed to
fabrication imperfections (roughness, inhomogeneities) of the membrane.
The lineshape and the decrease of the linewidth with angle
are well reproduced by the model, the modulations of R$_{max}$ and T$_{max}$ corresponding to the
transparency and absorption bands of the material.


The resonance can lead to perfect optical extinction below the onset
of the first diffraction order for a non-lossy material \cite{Gomez_OE06},
with a huge enhancement of the field on the rods and between
them \cite{Laroche_OL07}. For a material with an absorption band,
strong absorption resonances were also predicted \cite{Laroche_PRB06}. In this context,
it is instructive to present the absorption spectra
(see Fig. \ref{comp_expe_analytical}(b)). We first consider the absorption spectrum for
a given incidence angle. It is computed as ${\rm A} = 1-{\rm R} -{\rm T}$, and includes light diffusion due
to inhomogeneities and residual roughness, which is supposed to be negligible. Below
the resonance wavelength, the curve is dominated by the contribution of the first-order
diffraction wave (about 10~\%).
Above the resonance wavelength, the absorption drops almost to zero.
It is worth mentioning that off-resonance (e.g. blue curves at 4.6~$\mu$m) a membrane with 
only 15~\% material is absorbing just 50 \% less than a full non-structured membrane
(3 and 7~\%, respectively). 

At resonance, the absorption features a maximum. Remarkably, when the resonance is matching the Si-H band ($\lambda \simeq 4.6~\rm{\mu m}$),
the absorption efficiency of the membrane is enhanced by more than one order of magnitude as compared to non-resonant absorption at the same wavelength, reaching 33~\%.
This value is comparable with the maximum value of 50~\%, predicted for the absorption of similar systems \cite{Laroche_PRB06}.
The absorption cross-sections can be defined as: $\sigma_a = D (1-T-R)$ \cite{Barnes_PRL08}, where $\sigma_a$ is
expressed in unit of length due to the translational symmetry in the $y$ direction.
The volume absorption coefficient $\alpha_v = \sigma_a / d^2$, defined as the cross-section per unit
particle volume \cite{BohrenHuffman}, reaches a value of $\alpha_v = 4~\rm{\mu m}^{-1}$.
It can be compared to the absorption coefficient of the unstructured membrane, defined
as $\exp(-\alpha d)=$R+T: $\alpha = 0.16~\rm{\mu m}^{-1}$. These results show that for a fixed volume
of material, the absorption enhancement factor in the nanorod array compared to the thin film case reaches 25. 
This behavior is completely reproduced by the model, as shown in the inset of Fig. \ref{comp_expe_analytical}(b). 

\begin{figure*}
\begin{center}
\includegraphics[width=\textwidth]{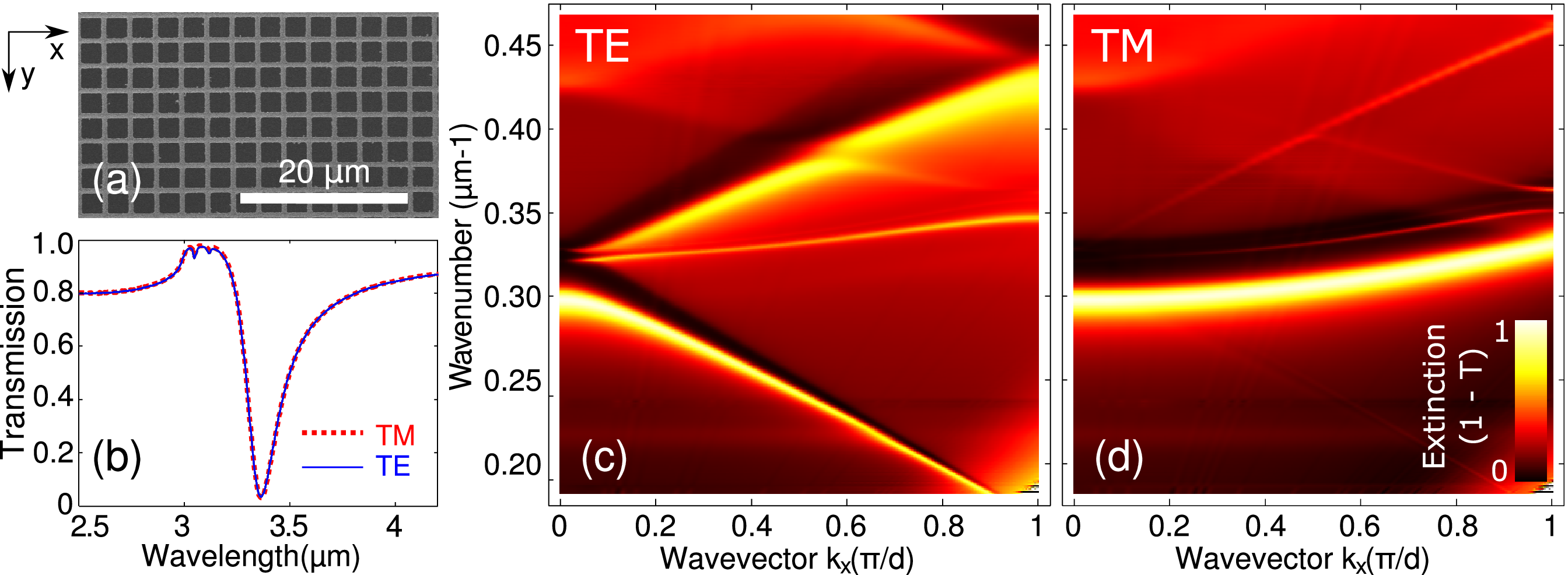}
\caption{
Free-standing dielectric membranes with 2D square design. {\bf a}, Scanning electron microscopy
image of the structure (period $D = 3 \: \rm \mu m$ and rod size $d = 500$ nm). {\bf b}, Transmission spectra acquired at normal incidence in 
both TE and TM polarizations and 
{\bf c-d}, Optical extinction dispersion relation for the two orthogonal polarizations.} \label{2D}
\end{center}
\end{figure*}

These properties can be fully extended to 2D arrays of nanorods, much more suited to practical applications. 
Indeed, the square 2D arrays allows improved robustness and insensitivity to polarization at normal incidence. 
Free-standing membranes with square 2D patterns were fabricated with the same geometrical parameters as the 1D analogue
(width of the rods 500 nm, grating period 3 $\mu$m), see Fig. \ref{2D}(a). 
Fig. \ref{2D}(b) shows the transmission spectra measured at normal incidence.
$97\%$ optical extinction is obtained for both polarizations (TE and TM). As previously,
Fano resonances are observed, at slightly larger wavelengths (3.36 $\mu$m) 
with even higher extinction efficiency than in 1D structures.
Actually, at normal incidence, the structure behaves as a superposition of two orthogonal 1D nanorod arrays.


At oblique incidence, the optical response is modified as compared with the 1D case.
The dispersion diagrams of the optical extinction are shown in Fig. \ref{2D}(c-d).
For the TE case, the two dispersion branches observed in the 1D grating are also present.
It originates from multiple scattering involving the excitation of the dipole moment
\textbf{$p_y$} of the y-axis nanorods.
An additional thin, flat band appears at 3.1~$\mu$m. It follows the Rayleigh anomalies
of the diffracted waves having the in-plane wavevectors
$\textbf{k}= (2\pi / \lambda)\sin \theta \ \textbf{x} \pm (2\pi / D) \ \textbf{y}$.
The dipole moments \textbf{$p_x$} of the x-axis nanorods are excited by the small
x-component of the electric field of these diffracted waves for $\theta \neq 0$ 
\cite{PhysRevB.79.165405, Sauvan08apl}. A similar, wider band is
observed for TM polarization. It is related to the dipole moments \textbf{$p_x$}
of the x-axis nanorods. 

\begin{figure*}
\begin{center}
\includegraphics[width=\columnwidth]{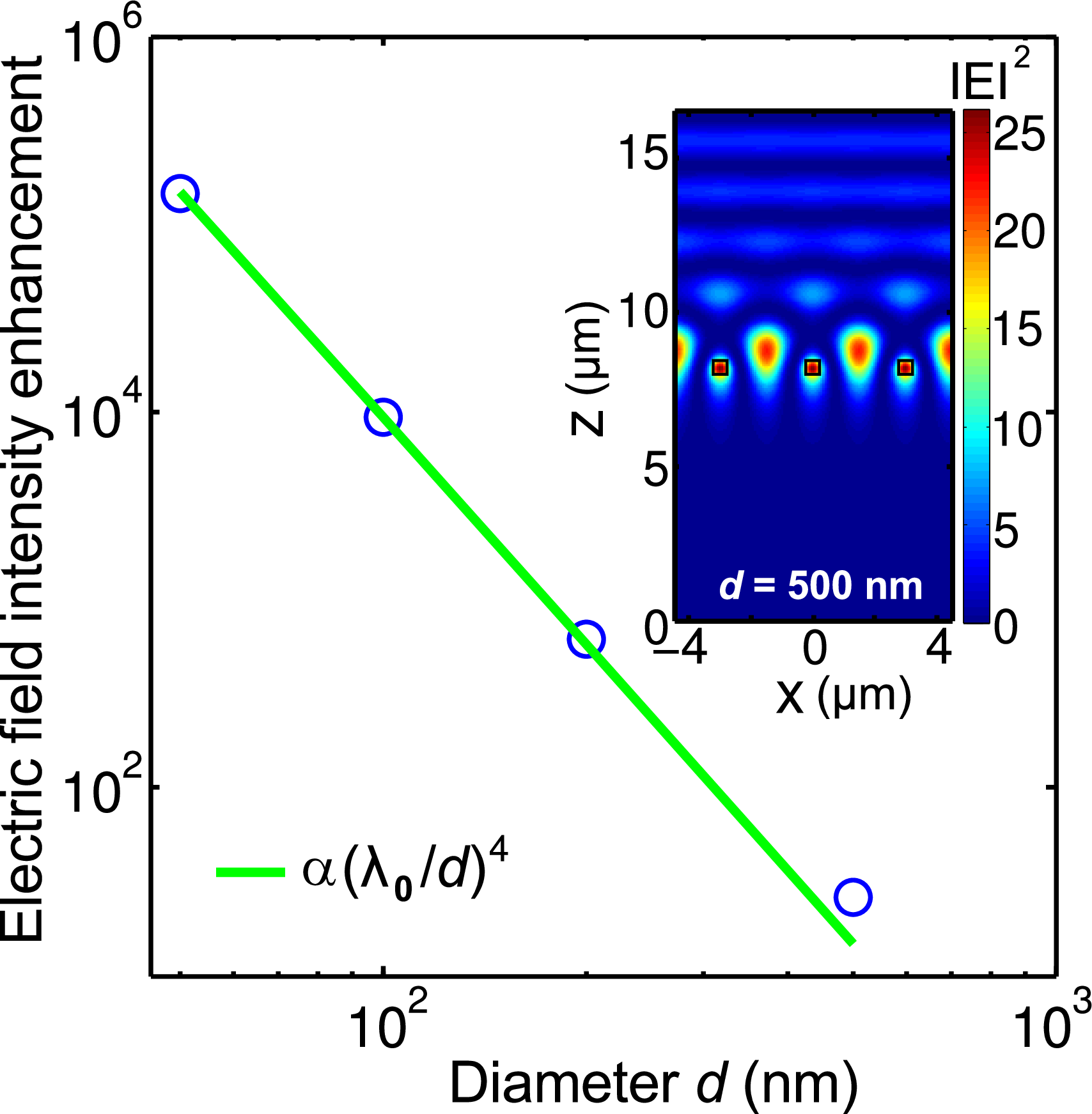}
\caption{
Electric-field enhancement in 1D nanorod arrays as a function of the diameter $d$, at resonance ($\theta=0^{\circ}$).
Circles: full electromagnetic calculation. Line: ($\lambda_0/d)^4$ law, where the resonance wavelength can be approximated by $\lambda_0 \simeq 3 \: \rm \mu m$ for small diameters.
Inset: Map of the electric-field intensity for $d$= 500 nm, normalized to the incident field. The position of the rods is shown with black squares.} \label{ID}
\end{center}
\end{figure*}

In relation with the sharp extinction resonances, the nanorod array exhibits very high field enhancements. The map of the electric field intensity, and the evolution of the field enhancement with the rod diameter, have been calculated by rigourous coupled-wave analysis and are reported in Fig.  \ref{ID}, for 1D arrays~\cite{Reticolo,Supl.Info}. It shows that the maximum of the electric field is located in the center of the nanorods. Importantly, as the nanorod diameter $d$ decreases, the electric-field intensity increases as $(\lambda/d)^4$ \cite{Laroche_OL07}. Field enhancements as large as $10^5$ are obtained for nanorod diameters of 50 nm. It opens the way to the conception of high-Q dielectric structures with very low volumes, despite the large surface area covered by the nanorod array. In view of applications, the influence of the finite size of the array on the mode volume and on the field-enhancement is an important issue that should be addressed. It is also important to note that the linewidth and the spectral position of the resonance can be easily tuned by varying the geometrical parameters of the nanorod array (period, diameter) or the external environment. The resonance is not restricted to the range of frequencies of plasmons, and can be shifted from the visible to the far-IR wavelength range.

In conclusion, we have described and measured the far-field response of free-standing, subwavelength
dielectric nanorod arrays. Extremely sharp peaks together with nearly total extinction was
experimentally demonstrated. When the rods are made of a lossy material, an absorption enhancement
factor of 25 is found. These singular properties originate from two important features: (i) scatterers are non-resonant and much smaller than the wavelength, and (ii) free-space interactions between the scatterers.
It is worth noticing the similarities between the coherent multiple scattering
mechanism evidenced in nanorod arrays, and the well-known Bragg diffraction arising in the
three-dimensional arrangement of a crystal lattice. Both phenomena are based on low cross-section
scatterers. However, the constructive interference of the Bragg diffraction mechanism involves
a large number of lattice planes. Surprisingly, it is shown here that nearly 100~\% of incident
photons interact with a single layer of sparse scatterers periodically arranged in a symmetrical
environment.
These properties could have an important impact for applications 
in surface enhanced Raman scattering \cite{Martin01}, fluorescence enhancement
\cite{Vahid06}, nonlinear optics \cite{Novotny07}, 
and coherent light emission \cite{Greffet02}.
Beyond, a wide variety of stop-band filters and selective mirrors can be achieved with
nanostructured membranes, with potential applications in multispectral imaging \cite{CollinFilters,HaidarAPL}
and optomechanics \cite{Kippenberg2008,Thompson2008}.
As an illustration, an efficient optomechanical interaction has been achieved recently with an unstructured membrane placed in a high-finesse cavity \cite{Thompson2008}. With increased reflectivity and drastic reduction of the mirror mass, free-standing nanorod arrays could offer strong improvement of the optomechanical coupling, and open new possibilities to probe the quantum regime of mechanical systems.

{\bf Acknowledgements}

This work was partially supported by the ANR project Metaphotonique and the PRF Metamat. We acknowledge J.J. Saenz,
S. Albaladejo, S. Maine, S. Vassant, J. Jaeck and S. Rommelu\`ere for fruitful discussions.

\newpage

\newpage

\newpage

\newpage

\newpage

\end{document}